\newcommand{\CH}{\hbox{{$\cal H$}}}

\newcommand{\R}{\mathbb{R}}
\newcommand{\N}{\mathbb{N}}
\newcommand{\C}{\mathbb{C}}

\newcommand{\note}[1]{}

\newcommand{\extd}{{\rm d}}

\newcommand{\isom}{{\cong}}
\newcommand{\eps}{{\epsilon}}
\newcommand{\tens}{\mathop{\otimes}}
\newcommand{\id}{{\rm id}}

\renewcommand{\>}{\rangle}
\newcommand{\del}{\partial}

\newcommand{\und}[1]{{\underline {#1}}}

\newcommand{\lcross}{{>\!\!\!\triangleleft}}

\newcommand{\eqn}[2]{\begin{equation}#2\label{#1}\end{equation}}

\documentclass[12pt,A4]{article}
\usepackage{amssymb,amsmath}
\begin{document}\baselineskip 18pt

{\ }\qquad \hskip 4.3in \vspace{.2in}

\begin{center} {\Large  ON THE  FOCK SPACE FOR NONRELATIVISTIC ANYON FIELDS AND BRAIDED TENSOR PRODUCTS}
\\ \baselineskip 13pt{\ }\\
{\ }\\   Gerald A. Goldin\footnote{Professor} \\ {\ }\\
Departments
of Mathematics and Physics\\
Rutgers University\\
New Brunswick, NJ 08903, USA\\
\smallskip
{\it gagoldin@dimacs.rutgers.edu\/}\\
{\ }\\
+\\{\ }\\ Shahn Majid\footnote{Royal Society University Research
Fellow}
\\ {\ }\\ School of
Mathematical Sciences\\
Queen Mary, University of London, Mile End Rd\\
London E1 4NS, UK\\
\smallskip
{\it s.majid@qmw.uc.uk\/}\\
\end{center}
\begin{center}

\end{center}

\begin{quote}\baselineskip 14pt
\noindent{\bf Abstract} We realize the physical $N$-anyon  Hilbert
spaces, introduced previously via unitary representations of the
group of diffeomorphisms of the plane, as $N$-fold
braided-symmetric tensor products of the $1$-particle Hilbert
space. This perspective provides a convenient Fock space
construction for nonrelativistic anyon quantum fields along the
more usual lines of boson and fermion fields, but in a braided
category, and clarifies how discrete (lattice) anyon fields relate
to anyon fields in the continuum. We also see how essential
physical information is encoded. In particular, we show how the
algebraic structure of the anyonic Fock space leads to a natural
anyonic exclusion principle related to intermediate occupation
number statistics, and obtain the partition function for an
idealized gas of fixed anyonic vortices.
\end{quote}

\baselineskip 18pt
\section{Introduction}

Anyons are particles or excitations in two-dimensional space that
obey exchange statistics interpolating those of bosons and
fermions. When two identical anyons are exchanged without
coincidence along a continuous path in the plane, their relative
winding number $m$ (the net number of counterclockwise exchanges)
is well-defined, depending only on the homotopy class of the path
implementing the exchange. The quantum-mechanical wave function
then acquires a relative phase $e^{\,i m\theta}$, where $\theta$
is a real fixed parameter between 0 and $2\pi$. When $\theta=0$ we
have bosons, and $\theta=\pi$ corresponds to fermions.

The possibility of such intermediate statistics was suggested by
Leinaas and Myrheim \cite{LM} and confirmed by Goldin, Menikoff
and Sharp \cite{GMS}, who derived the quantum theory rigorously
from representations of local nonrelativistic current algebra and
the corresponding diffeomorphism group. They obtained the anyonic
shifts in angular momentum and energy spectra, and made
connections with configuration space topology and the physics of
charged particles circling regions of magnetic flux. The term
``anyon'' was subsequently introduced by Wilczek \cite{Wil}, who
proposed a model for such objects based on
charged-particle/flux-tube composites and suggested their
association with fractional spin in two dimensions. The idea found
some immediate applications to surface phenomena and related areas
of physics. \cite{Laughlin1983, Halperin1984}

In \cite{GS1983} the {\it braid group\/} $B_N$ was identified as
the group whose one-dimensional representations describe the
anyonic wave function symmetry. A more extensive discussion of the
braid group and anyon statistics followed in \cite{Wu}, where it
was argued that only the one-dimensional representations of $B_N$
should occur in quantum mechanics. However, as noted in
\cite{GMS1985}, the diffeomorphism group approach allows also the
possibility of quantum systems described by higher-dimensional
representations of $B_N$ (particles later termed ``plektons''). An
overview of that approach, which underlies the present article,
may be found in \cite{GS1991}.

The extensive development of these ideas that occurred during the
1980s and early 1990s, including their relation to Chern-Simons
quantum field theories, their application in describing the
integer and fractional quantum Hall effects, and their role in
describing possible mechanisms for superconductivity, are reviewed
in \cite{Wilczek1990} and \cite{Khare1997}.

Anyonic systems can also be associated with quantum groups and
$q$-deformations of classical Lie algebras
\cite{Ma:any,LerdaSciuto1993,FMS1994,FLS1994,LM1995}. In
\cite{GS1996} it was shown that creation and annihilation fields
for anyons could be constructed so as to intertwine the $N$-anyon
representations of the group of compactly supported
diffeomorphisms {\it Diff\/}$_c(\R^2)$ in the Hilbert space
$\CH_N$ of $N$-particle states. The assumption that these fields
transform consistently with the diffeomorphism group
representations dictates the form they should take; and the fact
that they obey $q$-commutation relations, where $q$ is the anyonic
phase shift, emerges as a consequence of this. In this work, the
spaces $\CH_N$ were constructed (for each $N$) using `topological
configurations' of $N$ points equipped with attached filaments
going out to infinity in the plane.

In the present article, we show how to construct such a theory
along lines more familiar from bosonic (respectively, fermionic)
algebras of canonical commutation (resp. anticommutation)
relations---i.e., CCR (resp. CAR) algebras---where the
$N$-particle Fock subspaces are built up by symmetrizing or
antisymmetrizing sets of 1-particle states. We shall effectively
`$q$-symmetrize' \cite{Mitra1995} using braided category
techniques coming out of quantum group theory \cite{Ma:book}; so
that \[ \CH_N = \CH_1\tens_s \CH_1\tens_s\cdots \tens_s \CH_1\]
($N$ times), where $\tens_s$ is the symmetrized tensor product
with respect to a certain symmetry $\Psi_0$. An apparent
difficulty in formulating anyonic field theory this way has been
the need to work in a {\it strictly\/} braided category, with a
braiding $\Psi$ defined by $q$. The fact that the braid group is
infinite would then seem to require an infinite sum of powers of
$q$. But it turns out that the associated operator $\Psi_0$ obeys
the condition $\Psi_0^2 = \id$. Thus it generates an action of the
symmetric group $S_N$, rather than the braid group $B_N$. In
effect, the nontrivial braid group representation and its inverse
conspire with each other to give us a braided tensor category that
is actually {\it symmetric\/} (up to sets of measure zero).

Moreover, in our construction the full anyonic Fock space
$S_{\Psi_0}(\CH_1)$, obtained as the direct sum of the different
spaces $\CH_1\tens_s\cdots\tens_s\CH_1$, is an algebra with
product $\tens_s$. This algebra generalizes the algebra of
functions on a linear space or superspace in the Bose or Fermi
cases, with its respective commutative or anticommutative product.
The annihilation and creation operators in our construction then
act by pointwise multiplication and (functional) differentiation
on this space, exactly in conformity with the usual functional
representation in field theory. The difference is that the anyonic
case only makes sense in a braided category, with a braiding
$\Psi$. In this category $S_{\Psi_0}(\CH_1)$ is the `coordinate
ring' of a braided group, with the braided coproduct $\und\Delta$
expressing addition \cite{Ma:any,Ma:book,Ma:csta,Ma:fre}. The use
here of {\em two} Yang-Baxter operators $\Psi$ and $\Psi_0$ is a
general feature of the theory of such `braided linear spaces', and
the strictness of the braiding $\Psi$ is essential for this.

As an important application, we show how the structure of
$S_{\Psi_0}(\CH_1)$ as the braided version of the coordinate
algebra of a linear space leads to an `anyonic exclusion
principle' when $q$ is a root of unity. Specifically, with
$q^{\,r}=1$, the creation operator $\psi^*(x)$ cannot occur in a
reduced Fock state more than $r-1$ times. This important physical
fact means that the relevant occupation number statistics is of
the nature of Gentile statistics \cite{Gentile}, and has some
similarities to other algebraic approaches to generalized
exclusion principles such as that proposed in \cite{Kibler}. The
exclusion also applies to states obtained from the smeared field;
i.e., $\psi^*(h)^{\,r}=0$ in our Hilbert space representation for
any test function $h$. Earlier articles have approached the
fractional exclusion statistics of a one-dimensional gas, or of
anyons in two space dimensions, from quite different perspectives
\cite{ Haldane1991,Wu1994,NayakWilczek1994,MurthyShankar1994,
CanrightJohnson1994,Medvedev1997}; see also \cite{Mitra1994}.

An outline of the present article is as follows. In Section~2 we
summarize the topological construction of the spaces $\CH_N$ of
$N$-anyon states, and write the corresponding current algebra,
diffeomorphism group, and anyon creation and annihilation field
representations. In Section~3 we carry out our construction using
braided symmetric tensor categories, and in Section~4 we show that
this indeed leads to the same result as the earlier topological
construction. In Section~5 we provide some algebraic consequences,
including the anyonic exclusion principle when $q$ is a root of
unity. Throughout we consider both the physical continuum case and
the discrete case, as the latter may be useful for lattice
versions of the theory or for anyon fields on a finite number of
points. The discrete case turns out to have a number of
instructive subtleties. Finally, in Section~6 we offer some
further comments of a conceptual nature.

\section{Topological state spaces $\CH_N$}

We begin by recalling the set-up for anyon field representations
proposed in \cite{GS1996}. For the nonrelativistic quantum theory
of identical particles in $n$ space dimensions, we are interested
in representations $\,\rho,\,J\,$ of the semidirect sum Lie
algebra $\,C^\infty_c(\R^n) \lcross \,\, {\rm vect}_c(\R^n)$,
where $\rho$ and $J$ are self-adjoint operator-valued
distributions describing the mass and momentum densities
respectively; i.e.,
\[ \rho\,(f)=\int \rho\,(x)f(x)\extd^n x,\quad J(v)
=\int J(x)\cdot v(x)\extd^n x,\]
where the test function $f \in C^\infty_c(\R^n)$ is a
compactly-supported real-valued $C^\infty$ function on $\R^n$, and
$v \in {\rm vect}_c(\R^n)$ is a compactly-supported (tangent)
vector field on $\R^n$. Then the well-known current algebra
\begin{equation}
[\rho\,(f), \rho\,(g)] = 0,\quad[\rho\,(f),J(v)]=i\hbar
\rho\,(\nabla_v f),\quad [J(v),J(w)]=-i\hbar J([v,w])
\label{currentalg}
\end{equation}
represents the bracket in the Lie algebra, where $[v,w]$ is the
usual Lie bracket of vector fields. The group-level version is
based on the natural semidirect product of the group of
compactly-supported functions under addition, with the group of
compactly-supported diffeomorphisms of $\R^n$ under composition:
$G = C^\infty_c(\R^n)\,\lcross\,\, {\rm Diff}_c(\R^n)$, with
$(f,\phi)\cdot(g,\psi)=(f+g\circ\phi,\psi\circ\phi)$. Then in a
continuous unitary representation of $G$, we can write the
1-parameter subgroups $U(f)$ and $V(\phi_t)$, where $\phi_t$ is
the flow on $\R^n$ generated by $v$. Under appropriate conditions,
the self-adjoint generators defined from $U(f) = e^{\,i
\rho\,(f)}$ and $V(\phi_t)=e^{\,i (t / \hbar) J(v)}$ represent the
current algebra. The idea is that different physical systems in
quantum mechanics should correspond to different (unitarily
inequivalent) irreducible representations of $G$.

In particular consider a family $\CH_N$ of Hilbert spaces, where
$N\in \N$, along with annihilation operators $\psi\,(h)$ and
creation operators $\psi^*(h)$, where the test functions $h$
belong to a domain in $\CH_1$. Thus
\[ \psi(h):\CH_{N+1}\to \CH_N,\quad \psi^*(h):\CH_N\to
\CH_{N+1}.\]
Suppose we have representations $U_N,V_N$ in $\CH_N$ of the group
$G = C^\infty_c(\R^n)\,\lcross\,\, {\rm Diff}_c(\R^n)$, for each
$N$, intertwined by $\psi$ and $\psi^*$ in such a way that
\begin{equation}
U_{N+1}(f)\psi^*(h)=\psi^*(U_1(f)h)U_N(f),\quad
V_{N+1}(\phi)\psi^*(h)=\psi^*(V_1(\phi)h)V_N(\phi).
\label{intertwinegroup}
\end{equation}
Then the $U_N, V_N\,\,(N = 1,2,3, \dots)$ are interpreted as a
{\it hierarchy\/} of representations of $G$ describing systems of
$N$ particles (or quantum excitations) of the species created and
annihilated by the field operators. At the level of the algebra,
the corresponding requirements are
\begin{equation}
[\,\rho(f),\psi^*(h)]=\psi^*(\rho_1(f)h),\quad
[J(v),\psi^*(h)]=\psi^*(J_1(v)h). \label{intertwinealg}
\end{equation}
Here $\psi^*$ is the adjoint of $\psi$; but we note that later,
when we consider the discrete anyonic case, that will be modified.

When $n = 2$, one has the possibility in this general framework of
anyonic representations of $G$, and corresponding fields
satisfying Eqs. (\ref{intertwinegroup}) and (\ref{intertwinealg}).
Then the representation $U_N,\,V_N$ of $G$ acts in the Hilbert
space
\[ \CH_N=L^2_{B_N}(\tilde\Delta_N),\]
defined as follows. The configuration space $\Delta_N$ for $N$
identical anyons is the space of (unordered) $N$-point subsets of
$\R^2$; thus $\gamma \in \Delta_N$ is given by
$\gamma=\{x_1,\dots, x_N\} \subset \R^2$. The fundamental group
$\pi_1(\Delta_N)$ is $B_N$, the braid group on $N$-strands. We
denote by $\tilde\Delta_N$ the universal covering space of
$\Delta_N$, which has infinitely many sheets; for $\tilde\gamma
\in \tilde\Delta_N$, we have the projection map $p: \tilde\gamma
\to \gamma$. The braid group then acts on $\tilde\Delta_N$;
writing this action as $\tilde\gamma \to \tilde\gamma \cdot b$ for
$b \in B_N$, we have $p\,(\tilde\gamma \cdot b) =
p\,(\tilde\gamma)$. The elements of $\CH_N$ are now wave functions
$\tilde\Phi$ on $\tilde\Delta_N$, taking values in an inner
product space $\mathcal{V}$ that carries a unitary representation
$T(b)$ of $B_N$. We shall consider only the case $\mathcal{V} =
\mathbb{C}$ (scalar-valued wave functions), and $1$-dimensional
representations of $B_N$. Such a representation is specified by
choosing a fixed phase $q = \exp i\theta$, and setting $T(b) =
q\,$ when $b$ is the crossing of one strand over another in a
forward (left over right) direction. The wave functions are
required to be {\it equivariant\/} under $T$, in the sense that
for all $b\in B_N$,
\[ \tilde\Phi(\tilde\gamma \cdot b)=T(b)\tilde\Phi(\tilde\gamma).\]
In other words, $\tilde\Phi \in \CH_N$ is an equivariant section
of a vector bundle over $\tilde\Delta_N$. When $\tilde\Phi_1$ and
$\tilde\Phi_2$ satisfy the same such equivariance condition, the
product
$\overline{\tilde\Phi_1(\tilde\gamma)}\tilde\Phi_2(\tilde\gamma)$
depends only on $\gamma = p\,(\tilde\gamma)$, and not on the
particular choice of $\tilde\gamma$ within $p^{-1}(\gamma)$ at
which $\tilde\Phi_1$ and $\tilde\Phi_2$ are evaluated. Thus we may
write the integral of this product with respect to a (local)
Lebesgue measure $dx_1\cdots dx_N$ on $\Delta_N$. Finally, we take
$\CH_N$ to consist of the square-integrable functions, so that for
any pair $\tilde\Phi_1, \tilde\Phi_2$,
\begin{equation}
\langle \tilde\Phi_1, \tilde\Phi_2 \rangle = \int_{\Delta_N}
\overline{\tilde\Phi_1(\tilde\gamma)}\tilde\Phi_2(\tilde\gamma)\,
dx_1 \cdots dx_N < \infty \label{innerproduct}
\end{equation}
defines the inner product of $\tilde\Phi_1$ with $\tilde\Phi_2$ in
$\CH_N$.

Given any diffeomorphism $\phi$ of $\R^2$, let $\phi$ act on
$\Delta_N$ in the obvious way. This action lifts to an action on
$\tilde\Delta_N$ compatible with the projection map; i.e., if
$p\,(\tilde\gamma) = \gamma$ then $p\,(\phi \tilde \gamma) = \phi
\gamma$. The $N$-anyon representation in $\CH_N$ is then defined
by
\begin{equation}
U_N(f)\tilde\Phi(\tilde\gamma)=e^{i\sum_{j=1}^N
f(x_j)}\tilde\Phi(\tilde \gamma),\quad
V_N(\phi)\tilde\Phi(\tilde\gamma)=\tilde\Phi(\phi\tilde\gamma)
\prod_{j=1}^N\sqrt{\mathcal{J}_\phi(x_j)}\,, \label{Nanyonreps}
\end{equation}
where $\mathcal{J}_\phi$ is the Jacobian of $\phi$. The factor of
$\prod_{j=1}^N\sqrt{\mathcal{J}_\phi(x_j)}$ means that $V_N$
transforms
$\overline{\tilde\Phi_1(\tilde\gamma)}\tilde\Phi_2(\tilde\gamma)\,
\prod_{j=1}^N dx_j\,\,$ to
$\,\,\overline{\tilde\Phi_1(\phi\tilde\gamma)}
\tilde\Phi_2(\phi\tilde\gamma)\, \prod_{j=1}^N
\mathcal{J}_\phi(x_j)dx_j$, and thus does not change the value of
the inner product.

One may construct these representations more explicitly as
follows. We describe an element of $\tilde\Delta_N$ by a set of
non-intersecting paths $\Gamma=\{\Gamma_1,\cdots,\Gamma_N\}$ in
the plane, extending from infinity in (let us say) the negative
$y$-direction, and terminating in the unordered set $\gamma$ of
$N$ distinct points in $\R^2$. Then $\tilde\Delta_N$ is the set of
homotopy classes of such $\Gamma$, with the projection map given
by mapping $\Gamma$ to the set of its endpoints. We shall call the
homotopy class of $\Gamma$ a ``topological configuration.''
Moreover, we can lift any configuration $\gamma\in \Delta_N$ (with
the exception of a measure zero set) to the element
$\Gamma_0^\gamma$ belonging to $\tilde\Delta_N$, given by taking
paths that go vertically downward (in the negative $y$-direction)
from each point in $\gamma$. This defines a sheet
$\Gamma_0(\Delta_N)\subset \tilde\Delta_N$ that we conventionally
associate with the identity element of $B_N$. Now diffeomorphisms
in $\mathrm{Diff}_c(\R^2)$ act as the identity at $y=-\infty$
since they are compactly supported, and so they lift from
$\Delta_N$ to act on the space of topological configurations.

Consider the subgroup $\mathrm{Diff}_c^{\,\gamma}(\R^2)$ of
diffeomorphisms that take the set of endpoints $\gamma$ of a fixed
topological configuration $\Gamma$ into itself. This is the
stability subgroup for the point $\gamma \in \Delta_N$. A
diffeomorphism $\phi \in \mathrm{Diff}_c^{\,\gamma}(\R^2)$ then
determines an element of $B_N$ by its action on
$\Gamma_0^{\,\gamma}$, denoted by $b = h_\gamma(\phi)$. Moreover,
given any topological configuration $\Gamma$ with endpoints
$\gamma$, we can obtain it by starting with $\Gamma_0^\gamma$
(with the same endpoints $\gamma$), and applying a diffeomorphism
$\phi \in \mathrm{Diff}_c^{\,\gamma}(\R^2)$, so that $\phi
\Gamma_0^\gamma = \Gamma$. Then $h_\gamma$ is a surjective
homomorphism, from $\mathrm{Diff}_c^{\,\gamma}(\R^2)$ onto $B_N$.
A topological configuration $\Gamma$ with end points $\gamma$ is
conventionally identified with the pair $(\gamma, b)$; when $b$ is
the identity element, we have $\Gamma^{\gamma}_0$. The
equivariance condition in this description becomes
\[
\tilde\Phi(\Gamma) = T(b)\tilde\Phi(\Gamma_0^\gamma).
\]
Thus it is enough to specify $\tilde\Phi$ on the sheet
$\Gamma_0(\Delta_N)$; the equivariance condition then defines it
almost everywhere in $\tilde \Delta_N$.

In this explicit realization, one next defines the creation and
annihilation fields intertwining the $N$-anyon representations
(\ref{Nanyonreps}) in accordance with Eqs. (\ref{intertwinegroup})
or (\ref{intertwinealg}). Given an $N$-point subset $\gamma
\subset \mathbb{R}^2$, and $x \in \mathbb{R}^2$, let us denote by
$\,\Gamma_x^{\gamma}\,$ the element of $\,\tilde \Delta_{N+1}\,$
that is obtained by {\it adjoining\/} to $\Gamma_0^{\gamma}$ an
additional path, terminating at $x$, that extends toward $y =
-\infty$ on the {\it left\/} of all the paths in
$\Gamma_0^{\gamma}$ (this modifies the convention in Ref.
\cite{GS1996}). Then we set
\[
(\psi(x)\tilde\Phi)(\Gamma_0^\gamma)=\tilde\Phi(\Gamma_x^\gamma).
\]
To write the adjoint field, let $\,\hat\gamma_j = \gamma -
\{x_j\}$, where $j$ refers to some indexing of the elements of
$\gamma$. The topological configuration $\Gamma_x^{\hat\gamma_j}$
then defines an element of $\tilde \Delta_N$, with the set of
terminal points $\hat\gamma_j \cup \{x\}$. Express
$\,\Gamma_x^{\hat\gamma_j}\,$ as $\phi \Gamma_0^{\hat\gamma_j \cup
\{x\}}$, and define $b_{\,x,\,j}=h_{\hat\gamma_j \,\cup
\{x\}}(\phi)$. Then
\[
(\psi^*(x)\tilde\Phi)(\Gamma_0^\gamma) =\sum_{j=1}^N\delta(x-x_j)
\tilde\Phi(\Gamma_0^{\hat\gamma_j})T^*(b_{\,x,\,j}).
\]

In this realization, one can recover the local current algebra
(\ref{currentalg}) by defining
\begin{equation}
\rho(x)=\psi^*(x)\psi(x),\quad J(x)={\hbar\over
2i}\left(\psi^*(x)(\nabla
\psi)(x)-(\nabla\psi^*)(x)\psi(x)\right), \label{rhoJdefs}
\end{equation}
where $\rho$ is the number density of anyons and $J$ is the
momentum density. Then also
\eqn{rhoJalg}{{} [\rho(f),\psi^*(h)]=\psi^*(fh),\quad
[J(v),\psi^*(h)]= {\hbar\over 2i}\psi^*(\nabla_v
h+\nabla\cdot(vh)),}
which are precisely Eqs. (\ref{intertwinealg}). Eqs.
(\ref{currentalg}), (\ref{intertwinealg}), and (\ref{rhoJdefs})
are the same in the anyonic case as in the usual Bose or Fermi
cases. But now we have, in place of the CCR or CAR algebras, the
following equal-time $q$-commutation relations \cite{GS1996}: in
the half-space $x^1 < y^1$, with $[A,B]_{\,q} = AB - qBA$ (where
the phase $q$ generates the representations $T(b)$ of $B_N$),
\eqn{psirel}{ [\psi(x),\psi(y)]_q=[\psi^*(x),\psi^*(y)]_q=0,\quad
[\psi(y),\psi^*(x)]_q=\delta(x-y). }
In the complementary half-space $x^1 > y^1$, $q$ must be replaced
by $\bar{q} = q^{-1}$. The choice of a half-space, like the
definition of the sheet $\Gamma_0(\Delta_N)$, is conventional and
has no physical consequence.

\section{Anyonic Fock space construction}

For the usual bosonic or fermionic representations, we have of
course a more conventional construction. Let $\CH=\CH_1=L^2(\R^n)$
be the space of $1$-particle states, and $\CH_N=\CH^{\tens_sN}$ be
a symmetrized or skew symmetrized tensor product. Then
$\,\psi^*(h)=h\tens_s\,$ and $\psi(h)$ is given by the interior
product, yielding the usual equal-time commutation or
anticommutation relations respectively. In this section we give
such a `Fock space' construction, more in line with the usual Bose
or Fermi cases but now with nontrivial $q$-statistics. We then
show in Section~4 that in the continuum case it is isomorphic to
the topological construction of Ref. \cite{GS1996} described
above. We shall use the machinery of braided linear spaces and
braided Weyl algebras as described in Refs. \cite{Ma:book},
\cite{Ma:csta}, \cite{Ma:fre}, and elsewhere.

Let us start with a construction that works for any totally
ordered space $(X,<)$. We shall initially take $X$ to be discrete.
Note that we do {\it not\/} assume here that $\psi^*$ is the
adjoint of $\psi$, although this turns out to be true in the
continuum case (see Section~6); a more complicated relation holds
between $\psi$ and $\psi^*$ in the discrete case. Subsequently we
consider $X=\R^2$, with $x<y$ if $x^1<y^1$, or if $x^1 = y^1$ then
$x^2<y^2$.

Let $\CH$ denote a space of functions on $X$, and $\{\delta_x\}$ a
basis of $\delta$-functions. We also define the functions
\[ \eps_0(x,y)=\begin{cases} 1&{\rm if}\ x<y\\ 0& x=y\\ -1 &
x>y\end{cases},\quad \eps(x,y)=\begin{cases} 1&{\rm if}\ x\le y\\
-1 & x>y\end{cases}.\]  These are almost, but not quite, the same.
Even in the continuum case they are not quite the same, because of
the existence of distributions with support on the set $x = y$).
In what follows, we derive a version of the equal time
$q$-commutation relations (\ref{psirel}) as:
\[ \psi(x)\psi(y)=q^{\eps_0(x,y)}\psi(y)\psi(x),\quad
\psi^*(x)\psi^*(y)=q^{\eps_0(x,y)}\psi^*(y)\psi^*(x)
\]
\eqn{q-weyl}{ \psi(x)\psi^*(y)-q^{\eps(y,x)}\psi^*(y)\psi(x)
=\delta(x-y),\quad \forall\ x,y\in X,}
where as noted $\psi^*$ is not the same as the adjoint operator
$\psi^\dagger$. Eqs. (\ref{q-weyl}) are the refinement of Eqs.
(\ref{psirel}) that comes out of our Fock space construction, and
are consistent with the representations of the fields described
using the topological configurations above.

\subsection{Discrete version}

Let $(X,<)$ be a discrete totally ordered space, $\CH$ a space of
functions on $X$, and $\delta_x$ the function that is $1$ on $x$
and $0$ elsewhere in $X$. We define the generalized flip (or
braiding) operators $\Psi,\Psi_0:\CH\tens \CH \to \CH\tens \CH$ by
\eqn{Psidelta}{
\Psi_0(\delta_x\tens\delta_y)=q^{\eps_0(x,y)}\delta_y\tens\delta_x,\quad
\Psi(\delta_x\tens\delta_y)=q^{\eps(x,y)}\delta_y\tens\delta_x.}
These both obey the familiar braid or Yang-Baxter relations
\[
\Psi_{12}\Psi_{23}\Psi_{12}=\Psi_{23}\Psi_{12}\Psi_{23},
\]
where the numerical subscripts refer to the position in
$\CH^{\tens 3}$. They also satisfy the cross-compatibility
conditions
\eqn{compat}{(\Psi+\id)(\Psi_0-\id)=0,\,\,\,
(\Psi_0)_{12}\Psi_{23}\Psi_{12}=\Psi_{23}\Psi_{12}(\Psi_0)_{23},\,\,\,
\Psi_{12}\Psi_{23}(\Psi_0)_{12}=(\Psi_0)_{23}\Psi_{12}\Psi_{23}.}
In addition $\Psi_0^2=\id$, but this fact and the braid relations
for $\Psi_0$ are not essential here.

One can write these operators in an $R$-matrix form:
$\Psi(\delta_x\tens\delta_y)=\delta_b\tens\delta_a
R^{\,a}{}_x{}^b{}_y$ (summing over the repeated variables $a,b$),
with $R^{\,a}{}_x{}^b{}_y=\delta^a{}_x\delta^b{}_yq^{\eps(x,y)}$;
and similarly for $\Psi_0$, in terms of a matrix $R{\,'}$.
Associated to any such $R,R{\,'}$-matrices
\cite[Th.~10.2.1]{Ma:book} is a `braided linear space' which in
our case we denote $S_{\Psi_0}(\CH)$. It is defined as the
quadratic algebra generated by formal products of the
$\{\delta_x\}$ generators, modulo the relations
\eqn{deltarel}{ \delta_x\tens_s
\delta_y\,=\,\tens_s\,(\Psi_0(\delta_x\tens\delta_y))\,\equiv\,
\delta_y\tens_s \delta_x \,q^{\eps_0(x,y)},}
where we denote the product in $S_{\Psi_0}(\CH)$  by $\tens_s$.
The actual braiding $\Psi$ enters in the braided coproduct
$\und\Delta:S_{\Psi_0}(\CH)\to S_{\Psi_0}(\CH)\tens
S_{\Psi_0}(\CH)$, that makes $S_{\Psi_0}(\CH)$ a braided group or
a Hopf algebra in a braided category. On generators, this is just
the linear braiding given by $\und\Delta\,\delta_x=\delta_x\tens
1+1\tens \delta_x$, but it extends to products under $\tens_s$ via
$\Psi$; thus
\eqn{coproduct}{
\und\Delta(\delta_x\tens_s\delta_y)=(\delta_x\tens_s\delta_y)\tens
1+\delta_x\tens \delta_y+q^{\eps(x,y)}\delta_y\tens\delta_x+
1\tens(\delta_x\tens_s\delta_y).}
As explained in \cite{Ma:book}, for the theory of braided spaces
one needs not one but {\em two} operators. One of these operators
controls the `internal' noncommutativity of the algebra, and the
other controls the `external' noncommutativity or braid statistics
with other independent copies. In our case, the physics actually
dictates the use of $\Psi$ rather than $\Psi_0$ in the second
role; as only this choice in Eq. (\ref{coproduct}) correctly
reduces to the constant minus sign in the flip for fermions when
$q = - 1$.

Next, we define the operators $\psi^*_x$ and $\psi^x$ on
$S_{\Psi_0}(\CH)$, by left multiplication and braided
differentiation respectively. Let
$\,[N+1,\Psi]=\id+\Psi_{12}+\Psi_{12}\Psi_{23}+ \,.\,.\,.\,
+\Psi_{12}\cdots\Psi_{N,N+1}\,\,$ be the braided integer matrix
\cite{Ma:book} \cite{Ma:fre}, where for example $\Psi_{12}$
denotes $\Psi$ acting in the $(1,2)$ pair of copies of $\CH$.
Then:
\begin{equation}\label{hatx} \psi^*_x
(\delta_{x_1}\tens_s\cdots\tens_s\delta_{x_N})\,=\,\delta_x\tens_s
\delta_{x_1}\tens_s\cdots\tens_s\delta_{x_N}\,,
\end{equation}
and
\begin{eqnarray}\label{delx}
\psi^x(\delta_{x_1}\tens_s\cdots\tens_s\delta_{x_{N+1}})\,&=&\,
\delta_{y_2}\tens_s\cdots\tens_s
\delta_{y_{N+1}}\,[\,N+1,\Psi]^{\,x\,y_2\,\cdots\,\,y_{N+1}}_{\,x_1\,x_2\,\cdots
\,\,x_{N+1}}\nonumber\\
&&\nonumber\\
&=&\delta^x_{x_1}\delta_{x_2}\tens_s\cdots\tens_s\delta_{x_{N+1}}
\,\,+\,\,
q^{\eps(x_1,\,x)}\delta^x_{x_2}\delta_{x_1}\tens_s\delta_{x_3}
\tens_s\cdots\tens_s\delta_{x_{N+1}}\nonumber\\
&&\nonumber\\
&&\quad+\,\,\cdots\,\,+\,\,
q^{\eps(x_1,\,x)+\,\cdots\,+\,\eps(x_{N},\,x)}\delta^x_{x_{N+1}}
\delta_{x_1}\tens_s\cdots\tens_s\delta_{x_{N}}.
\end{eqnarray}
Clearly the braided derivative $\psi^x$ (or $\del^{\,x}$ in the
notation of \cite{Ma:book}) is given by the evaluation or interior
product pairing with $\CH$, but extended to the whole of
$S_{\Psi_0}(\CH)$ as a braided derivation. This is like a
super-derivation, but using $\Psi$ to braid $\psi^x$ past elements
of $\CH$. These operators of braided differentiation arise as
infinitesimal translations on the braided space, as expressed
through the linear braided coproduct. From the braided Leibniz
rule for these operators, one has easily the following relations
\cite{Ma:book}:
\eqn{reldelta}{\psi^*_x\psi^*_y=\psi^*_b
\psi^*_aR\,'{}^{\,a}{}_x{}^b{}_y =\psi^*_y\hat\psi^*_x
q^{\eps_0(x,y)},\quad
\psi^x\psi^y=R\,'{}^{\,x}{}_a{}^y{}_b\,\psi^b\psi^a=
q^{\eps_0(x,y)}\psi^y\psi^x,}
and
\eqn{weyldelta}{\delta^x_y=\psi^x \psi^*_y-\psi^*_a
R^{\,a}{}_y{}^x{}_b\psi^b =\psi^x\psi^*_y-q^{\eps(y,x)}\psi^*_y
\psi^x,}
just as in Eqs. (\ref{q-weyl}). Notice that the $\psi^x$ operators
again obey the algebra $S_{\Psi_0}(\CH)$, as do (more obviously)
the $\psi^*_x$, but now we have also the cross relation
(\ref{weyldelta}).

To sum up, the space $S_{\Psi_0}(\CH)$, which should be viewed
geometrically as the algebra of $\Psi_0$-symmetric functions (on
the dual space $\CH$), first decomposes as a vector space. We have
$S_{\Psi_0}(\CH)=\oplus_N \CH^{\tens_s N}$, where the components
$\CH\tens_s\cdots\tens_s\CH$ ($N$ copies) consist of
$\Psi_0$-symmetric functions of degree $N$. We shall study these
components more explicitly in the next section. Secondly, the same
algebra acts by `pointwise multiplication' and `infinitesimal
translation' on itself, generating a braided Weyl algebra and
satisfying the relations (\ref{q-weyl}) in our discrete setting.
This construction generalizes both the usual CCR and CAR algebras,
along with their usual realizations on symmetric or antisymmetric
product algebras. All of this is an easy case of the general
theory of braided spaces and braided derivatives \cite{Ma:fre},
for our particular $R,R\,'$-matrices.

Note also that although we have carefully distinguished $\delta_x$
(the basis of the space $\CH$ generating $S_{\Psi_0}(\CH)$) from
the operators $\psi^*_x$ and $\psi^x$ that act on
$S_{\Psi_0}(\CH)$, we can also identify $\CH$ with the space
generated by the $\psi^*_x$ acting on $1$ by multiplication. Thus
we can identify the whole ``acted-upon'' copy of $S_{\Psi_0}(\CH)$
with the copy of $S_{\Psi_0}(\CH)$ generated by the $\psi^*_x$.
The vacuum state is $|\,0\,\>\,=\,1$. From the geometrical point
of view, the $\psi^x$ act by braided differentiation. From the
``Fock space'' point of view, they act via the commutation
relations and the condition that $\psi^x|\,0\,\>\,=\,0$.

Finally, let us specialize $X$ to a finite ordered set, writing
$X=\{1,\cdots,n\}$. Then using Eqs. (\ref{delx}), we have the
following results for $S_{\Psi_0}(\CH)$ and the corresponding
braided Weyl algebra:
\[ \psi^*_i\psi^*_j=q\psi^*_j\psi^*_i,\quad\psi^i\psi^j=q\psi^j\psi^i,
\quad \psi^i\psi^*_j=q^{-1}\psi^*_j\psi^i,\quad
\psi^j\psi^*_i=q\psi^*_i\psi^j,\quad (\,\forall \,i<j\,)\]
\begin{equation}
\psi^i\psi^*_i-q\psi^*_i \psi^i=1\,,\quad (\,\forall \,i\,)
\label{finitesetbrackets}
\end{equation}
where
\[
 \psi^*_i\,|\,m_1,\cdots,m_n\>
\,=\,q^{\,-\sum_{j<i}m_j}\,|\,m_1,\cdots,m_i+1,\cdots,m_n\>\,, \]
\eqn{psii}{ \psi^i\,|\,m_1,\cdots,m_n\>\,= \,q^{\,\sum_{j<i}m_j}
\,[m_i;q]\,|\,m_1,\cdots,m_i-1,\cdots,m_n\>\,.}
Here
\[ |\,m_1,\cdots,m_n\>\,=\,(\psi^*_1)^{m_1}\,\cdots\,(\psi^*_n)^{m_n}\,|\,0\>\,,\]
while
\[[m;q]\,=\,1+q+\,\cdots\,+q^{m-1}\,=\,(1-q^m)/(1-q)\]
is a `$q$-integer'. These equations yield, for example, the
``density operator'' as:
\eqn{rhoi}{\rho_i\,=\,\psi^*_i\psi^i,\quad
\rho_i|\,m_1,\cdots,m_n\>\,=\,[m_i;q]\,|\,m_1,\cdots,m_n\>.}
We note again that the final equation of (\ref{finitesetbrackets})
implies $\psi^*_i$ cannot be the adjoint of $\psi^i$ unless $q$ is
real. In general the $q$-integers in Eq. (\ref{rhoi}) are complex,
and the operator $\rho_i$ is not self-adjoint.

Although we are taking the above ``quantum-mechanical harmonic
oscillator'' point of view, the same mathematical structures can
also be viewed as spacetime position and momentum generators. Then
we would denote the generators $\delta_i$ by $X^i$, and regard
them as the coordinates of a noncommutative spacetime, with the
usual ``$q$-plane'' relations, $\,\,X_iX_j=qX_jX_i$ for $i<j$.
Similarly, we would write $\psi^*_i=\hat X_i$ for the operation of
left-multiplication by $X_i$, and $\psi^i=\del^{\, i}$ for the
braided differentiation
\[
\del^{\, i}(X_1^{m_1}\cdots X_n^{m_n})= q^{\,\sum_{j<i}m_j}\,
[\,m_i;q]\,X_1^{m_1}\,\cdots \,X_i^{m_i-1}\,\cdots
\,X_n^{m_n}\,.\]
This is the ``geometrical'' point of view to which we alluded
above. Here $\del^{\, i}$ acts on the $X_i$ variable by
$q$-differentiation, with an additional braiding $q$-factor as
$\del^{\, i}$ moves past the $X_j$, $j<i$, in order to reach the
$X_i$. Note that the $q$-plane is usually associated with more
complicated $R$-matrices related to quantum groups $SL_q(n)$ but
this is not the case here. Indeed, many different $R$-matrices can
give the same quantum plane.

\subsection{Functional version}

We next proceed to a functional version of the above, more
suitable for the continuum case. Clearly, on general functions
$h,g\in \CH$, the braiding is given by
\eqn{Psifg}{\Psi_0(h\tens g)(x,y)=g(x) h(y) q^{-\eps_0(x,y)},}
where $\Psi_0(h\tens g)=\int dx\,dy\, \Psi_0(h\tens g)(x,y)
\,\delta_x\tens\delta_y$ (the integrals here and below should be
interpreted as sums in the discrete case). Next, a basis for
$\,\CH\tens_s\cdots\tens_s \CH\,$ is given by ``normal ordering'';
i.e., we choose $\{\delta_{x_1}\tens_s\cdots\tens_s \delta_{x_N}|\
x_1\le x_2\le\cdots\le x_N\}$. We let the coefficients in this
basis be partially defined functions $f(x_1,x_2,\cdots,x_N)$ so
that
\[
f=\int_{x_1\le \cdots \le \,x_N} dx_1\cdots
dx_N\,f(x_1,x_2,\cdots,x_N)\,\delta_{x_1}\tens_s\cdots\tens_s\delta_{x_N}.\]

Then, defining $\psi^*(h)=\int\,dx\,h(x)\,\hat\delta_x$, we
compute:
\begin{eqnarray*}
\psi^*(h)f&=&\int_{x\,\le \,x_1\le \cdots
\le x_N}\,dx \,dx_1 \cdots dx_N\,h(x)f(x_1,\cdots,x_N)
\,\delta_x\tens_s\,\delta_{x_1}\tens_s\cdots\tens_s x_{x_N}\\
&&+\,q^{-1}\int_{x_1< x\,\le x_2\cdots\le x_N} \,dx_1 \,dx \cdots
dx_N h(x)f(x_1,\cdots,x_N)\,\delta_{x_1}\tens_s\delta_x\tens_s
\delta_{x_2}\tens_s\cdots\tens\delta_{x_N}\\
&&+\cdots+q^{-N}\int_{x_1\le\cdots\le x_N<\,x} \,dx_1 \cdots
dx_N\,dx\, h(x)f(x_1,\cdots,x_N)\,\delta_{x_1}\tens_s\cdots\tens_s
\delta_{x_N}\tens_s \delta_x\,,\end{eqnarray*}
\begin{equation}
\label{sumforpsistarhf}
\end{equation}
where we have used the relations (\ref{deltarel}). After
relabeling, we read off the coefficients as
\begin{eqnarray}\label{psi*h}
(\psi^*(h)f)(x_1,\cdots,x_{N+1})&=& h(x_1)f(x_2,\cdots,x_{N+1})\nonumber\\
&& \kern-100pt +q^{-1}h(x_2)f(x_1,x_3,\cdots,x_{N+1})+\cdots
+q^{-N}h(x_{N+1})f(x_1,\cdots,x_N),\end{eqnarray}
for distinct ``normally ordered'' arguments. This is our
representation of $\psi^*$ in the continuum case.

Similarly, we linearly extend the definition (\ref{delx}) of
$\del^{\,x}$, doing in each case one of the integrals and
relabeling the integration variables $x_1,\cdots,x_N$. This gives
\begin{eqnarray*}\psi(x)f&=& \int_{x\le x_1 \cdots\le
x_{N}}\,dx_1 \cdots
dx_N\,f(x,x_1,\cdots,x_{N})\delta_{x_1}\tens_s\cdots\tens_s
\delta_{N}\\ && +\,q\,\int_{x_1\le x\le x_2\le\cdots\le
x_{N}}\,dx_1 \cdots dx_N\,f(x_1,x,x_2,\cdots,x_N)
\delta_{x_1}\tens_s\cdots\tens_s \delta_{N}\\
&&+\cdots+\,q^N\,\int_{x_1\le\cdots\le x_N\le x}\,dx_1 \cdots
dx_N\,f(x_1,\cdots,x_N,x) \delta_{x_1}\tens_s\cdots\tens_s
\delta_{N},\end{eqnarray*}
which yields \eqn{psih}{(\psi(x)f)(x_1,\cdots,x_N)=q^m
f(x_1,\cdots,x_m,x,x_{m+1},\cdots,x_N)\,\quad {\rm for}\quad x_m<
x< x_{m+1}\,,}
with distinct, ``normally ordered'' $x_1,\cdots,x_N$ (using the
usual conventions for $m=0$ or $m = N$). This is our
representation of $\psi$ in the continuum case.

Finally, if the basis elements are taken in a different order, the
corresponding coefficients are related through Eq.
(\ref{deltarel}); e.g., we have
\[ f(x_2,x_1,\cdots,x_N)=q^{\eps_0(x_1,x_2)}f(x_1,x_2,\cdots,x_N),\]
and so forth. In this way, any coefficient function defined on
``normally ordered'' arguments extends uniquely to an element of
$\CH^{\tens N}$ obeying such relations.

Hence, proceeding now in the continuum case, we can characterize
$\CH\tens_s\cdots\tens_s \CH$ as the subspace of
`$\Psi_0$-symmetric' functions:
\eqn{Htens}{ \CH\tens_s\cdots\tens_s \CH=\{f\in \CH^{\tens N} |\
f(\sigma(x_1,\cdots ,x_N))=q^{\,\ell\,(\sigma)}f(x_1,\cdots,
x_N),\ \forall \, \sigma \in S_N,\ x_1<\cdots<x_N \}}
where $\ell\,(\sigma)$ is the length of the permutation $\sigma$.
It suffices to impose the symmetrization condition here almost
everywhere. For example,
\[ \CH\tens_s \CH=\{f|\ f(y,x)=qf(x,y),\quad \forall \,x<y \,\},\]
and
\[
\CH\tens_s\CH\tens_s \CH = \{f|\ f(y,x,z)=f(x,z,y)=qf(x,y,z),
\]
\[
f(z,x,y)=f(y,z,x)=q^2f(x,y,z),\,\, f(z,y,x)=q^3f(x,y,z),\,\,
\forall \,x<y<z\,\}.\]
Making use of the subspace description (\ref{Htens}), there is a
corresponding formula for the action of $\psi^*(h)$ that involves
factors of $q^\eps$. For example, letting $\psi^*(h)$ act on a
function $f \in \CH\tens_s \CH$, we have
\begin{eqnarray}\label{psistarh}
(\psi^*(h)f)(x,y,z)
&=&h(x)f(y,z)+q^{\eps(y,x)}h(y)f(x,z)+q^{\eps(z,x)+\eps(z,y)}h(z)f(x,y).
\end{eqnarray}
The general formula for degree $N$ is
\eqn{psi*tens}{
\psi^*(h)f=(\id\,+\,\Psi_{12}\,+\,\Psi_{23}\Psi_{12}
\,+\,\,.\,.\,.\,\,+\,\Psi_{N,\,N+1}\cdots\Psi_{23}\Psi_{12})(h\tens
f),}
which up to a normalization is just $h\tens f$ followed by total
$\Psi$-symmetrization (given the assumed $\Psi$-symmetry of $f$).
Similarly, the formula for $\psi(x)$ may be written in the
subspace description. It comes out simply as the interior product
\eqn{psitens}{ \psi(x)f=f(x,\ )\,,}
where we evaluate the first argument of $f$ at $x$. These are the
field operators when we work with $N$-particle wave functions as
symmetrized functions in $N$ variables, as in the usual Bose or
Fermi Fock spaces.

When there are products of fields, we will typically smear at
least some of the variables with test functions to make sense of
the distributions. Thus the $q$-commutation relations can be
written as
\[ \psi(x)\psi(h)=\psi(\theta_0{}_x h)\psi(x),\quad
\psi^*(x)\psi^*(h)=\psi^*(\theta_0{}_xh)\psi^*(x)\,\]
\eqn{q-weylh}{\psi(h)\psi^*(x)=\psi^*(x)\psi(\theta_x h)+h(x)\,,}
%
%
for $x \in X$ and $h$ in a dense domain of $\CH$, where for each
fixed $x$ we define
\[ \theta_0{}_x(y)=q^{\eps_0(x,y)},\quad \theta_x(y)=q^{\eps(x,y)}\,\]
as functions of $y$ (having modulus $1$ when $q$ is a phase), and
where $\theta_0{}_x h$ and $\theta_x h$ refer to the action on
$\CH$ given by pointwise multiplication. The second equation of
(\ref{q-weylh}) can also be written
\[ \psi(x)\psi^*(h )=\psi^*(\bar\theta_x h)\psi(x)+h(x)\,,\]
where $\bar\theta_x(y)=q^{\eps(y,x)}$.

As before, essentially the same algebra $S_{\Psi_0}(\CH)$ leads
both to the Hilbert space of the system with components
$\CH^{\,\tens_s N}$ as above (ignoring its algebra structure), and
to the generalized field algebra of the $\psi^*(h)$ together with
another copy for the $\psi(h)$ as an algebra of operator-valued
distributions obeying (\ref{q-weyl}) and represented according to
(\ref{psi*h})-(\ref{psih}). The Fock space can also be viewed as
generated by the operators $\psi^*(h)$ for a sufficient set of
test functions $h$, acting repeatedly on $|\,0\,\>$.

We also have a geometrical picture as at the end of the last
section, with $\psi(h)$ acting now by `braided functional
differentiation'.

\subsection{Unitarity considerations}

Until now we have worked rather generally, and have taken $q$ to
be arbitrary. We now work specifically over $\C$, and take $q$ to
be a phase. Addressing first the continuum case, we consider the
$L^2$ inner product on $\CH$, and verify that all our operator
constructions are suitably self-adjoint.
%
%

First we see that $\Psi_0$ is self-adjoint, since
\begin{eqnarray*}(a\tens b,\Psi_0(h\tens g))&=&\int\int\,dx\,
dy\, \bar a(x)\bar b(y)
g(x)h(y)q^{-\eps_0(x,y)}\\
&&=\int\int\,dx\, dy\,  \overline{\Psi_0(a\tens
b)(y,x)}h(y)g(x)=(\Psi_0(a\tens b),h\tens g).\end{eqnarray*}

Next we make use of the fact that the spaces
$\,\CH\tens_s\cdots\tens_s \CH\,$ have $L^2$ inner products, when
we write their elements as functions $f$ defined on the
fundamental domain of ``normally ordered'' coordinates. We perform
the integration over this domain. Thus:
\begin{eqnarray}
(g,\psi^*(h)f)&=&\int_{x_1\le\cdots\le x_{N+1}}
\,dx_1 \cdots dx_{N+1}\,\,\bar g(x_1,\cdots,x_{N+1})\cdot \nonumber\\
& & \quad \quad \quad \quad \quad \quad \quad \cdot
\sum_{m=0}^{N+1}
q^{-m}h(x_{m+1})f(x_1,\cdots,\hat{x}_{m+1},\cdots,x_{N+1})\nonumber\\
&=&\sum_{m=0}^N\,\int_{x_1\le \cdots\le x_N}\,dx_1 \cdots
dx_{N}\,\,\cdot\nonumber\\
& & \quad \quad \quad \cdot \,\overline{\int_{x_m}^{x_{m+1}}\,dx\,
q^m \bar h(x)
g(x_1,\cdots,x_m,x,x_{m+1},\cdots,x_N)}f(x_1,\cdots,x_N)\nonumber\\
& &\nonumber\\
&=&(\psi(\bar h)g,f)\,,\label{innerprodcont}\end{eqnarray}
where we have used the previous results for $\psi^*(h)$ and
$\psi(x)$, interpreted $\psi(\bar h)$ as the integral of $\psi(x)$
times $\bar h$, and relabeled the $x_i$ in the calculation. The
cases $m=0$ and $m=N$ are understood in the obvious way (the
integration is then taken to $\pm\infty$). In the summations for
$\psi^*(h)f$ (see Eq. (\ref{sumforpsistarhf})) some of the
inequalities in the region of integration are strict, but we are
permitted to ignore this distinction in the present continuum
case.

Finally, in our subspace description of $\CH\tens_s\cdots\tens_s
\CH$ we would like to define the inner product in terms of that on
$\CH^{\tens N}$. Indeed, for $f,g$ extended to $\Psi_0$-symmetric
elements of $\CH^{\tens N}$, we have
\[ (g,f)_{\CH^{\tens N}}=\sum_{\sigma\,\in \,S_N}\int_{x_1\le\cdots\le
x_N}\,dx_1 \cdots dx_N\,\bar
g(\sigma(x_1,\cdots,x_N))f(\sigma(x_1,\cdots,x_N))=N!\,(f,g).\]
That is, the natural inner product with respect to which $\psi$
and $\psi^*$ are mutually adjoint is $N!^{-1}$ times the usual
tensor product inner product. Alternatively, if one wished to use
the usual tensor product inner product, then one should work with
$(N+1)^{-1}\psi^*(h)$. From Eq. (\ref{psi*tens}) we see that this
would then be a true averaging over the symmetric group, and would
correspond to the usual normalization in the Bose and Fermi cases.

For the continuum case, we see that $\psi^*$ is the adjoint of
$\psi$ as in Ref. \cite{GS1996}. The discrete case requires us
perform summations instead of integrations, and to be careful
about the domains of summation. Thus, in the second term of Eq.
(\ref{sumforpsistarhf}), we have $x_1<x\le x_2 \le\,\cdots \, \le
x_N$; and similarly for the other terms. Then, as a correction to
Eq. (\ref{innerprodcont}), we have the following:
\begin{eqnarray}
(\psi(\bar h)g,f)&=&(g,\psi^*(h)f)\nonumber\\
&&\!\!\!\!\!\! +\!\!\!\!\!\!\! \sum_{x_1\le x_2 \le\,\cdots\,\le
x_N}\!\!\! \!\!\!\! \left(q^{-1}h(x_1)\bar g(x_1,x_1,x_2,\cdots
x_N)\, +\cdots +\, q^{-N} h(x_N) \bar
g(x_1,\cdots,x_{N-1},x_N,x_N)\right)\cdot\nonumber\\
&&\qquad\qquad \qquad\qquad \qquad\qquad \qquad\qquad
\qquad\qquad\qquad\qquad \cdot f(x_1,\cdots x_N)\,.
\end{eqnarray}
Writing $\psi(\bar h)^\dagger=\psi^*(h)+T^{\,*}_h$, where
$\psi(\bar h)^\dagger$ is the adjoint of $\psi(\bar h)$, we find
\[ T^{\,*}_h(\delta_{x_1}\tens_s\cdots\tens_s \delta_{x_N})
=\sum_{m=1}^Nq^{-m}h(x_m)\, \delta_{x_1}\tens_s
\cdots\tens_s\delta_{x_m}\tens_s \delta_{x_m}\tens_s\cdots\tens_s
\delta_{x_N}\]
where $\delta_{x_m}$ is duplicated on the right-hand side, and
where $x_1\le \cdots\le x_N$.

To proceed further, it is convenient (though not essential) to use
the simplified notation for the set $X=\{1,\cdots, n\}$. We then
find
\[ T_i^{\,*}((\delta_1)^{m_1}\cdots
(\delta_n)^{m_n})=q^{-\sum_{j\,<\,i}m_j}(q^{-1}+\cdots+q^{-m_i})(\delta_1^{m_1})
\cdots (\delta_i^{m_i+1})\cdots (\delta_n)^{m_n}\]
(where we have omitted the symbols $\tens_s$ for the symmetrized
tensor product). Hence
\[
T_i^{\,*}\,|\,m_1,\cdots,m_n\>
\,=\,q^{-1}q^{-\sum_{j\,<\,i}m_j}\,[\,m_i;q^{-1}\,]\,|\,m_1,\cdots,m_i+1,\cdots
m_n\>\,,\]
and
\[
{\psi^i}^\dagger\,|\,m_1,\cdots,m_n\>\,=\,
(1+q^{-1}[\,m_i;q^{-1}\,]\,)\psi_i^*\,|\,m_1,\cdots,m_n\>
\,=\,[\,m_i+1;q^{-1}\,]\,\psi_i^*\,|\,m_1,\cdots,m_n\>\,.\]
Noting that $\rho_{\,i}^\dagger$ has values which are the complex
conjugates of the values of the diagonal operator $\rho_{\,i}$ in
Eq. (\ref{rhoi}), we see that
\eqn{psi*dag}{
{\psi^i}^\dagger\,=\,\rho_{\,i}^\dagger\psi_i^*\,,\quad
\psi^i\,=\,(\psi_i^*)^\dagger \rho_{\,i}\,.}
It is worth remarking that
\eqn{psidagpsi}{
{\psi^i}^\dagger\psi^i\,|\,m_1,\cdots,m_n\>=\rho_{\,i}^\dagger\,
\rho_{\,i}=|\,[\,m_i;q\,]|^{\,2}\,|\,m_1,\cdots,m_n\>=
\frac{1-\cos\,m_i\,\theta}{ 1-\cos \theta}\,|\,m_1,\cdots,m_n\>}
if $q\,=\,e^{\,i\theta}$.

While the density operators $\rho_{\,i}$, are not self-adjoint in
the discrete case, they have some nice properties. For example
\[
[\rho_i,\psi^*_j]=0=[\rho_i,\psi^j],\quad(\forall\, i\ne j)
\]
but
\eqn{rhopsi}{\rho_i\psi^*_i-q\psi^*_i\rho_i=\psi^*_i,\quad
\psi^i\rho_i-q\rho_i\psi^i=\psi^i\,,}
with similar relations for $\rho_i^\dagger$; also
$[\rho_i,\rho_i^\dagger]=0$. The results differ from the continuum
case, where we already know that Eq. (\ref{rhoJalg}) holds with
the usual commutator, not the $q$-commutator of Eq.
(\ref{rhopsi}).

The same conclusions apply for any discrete set $X$. The origin of
the differences from the corresponding equations in the continuum
case is that in the discrete case, we have treated the
$\delta$-function in Eq. (\ref{q-weyl}) as a Kronecker-$\delta$
with values $0$ and $1$. This makes it the ``same size'' as the
$\delta$-functions arising in the combinatorics of the summations.
Such a treatment leads to elegant $q$-deformation formulas, as
obtained above. For a $\Bbb Z^n$-lattice theory that correctly
converges in the zero-spacing limit to the continuum theory, one
could use $\delta_{x,y}/\Delta^n$ in the right hand side of Eq.
(\ref{q-weyl}), where $\Delta$ is the lattice spacing.

\section{Identification with the topological picture}

It remains to identify the Hilbert spaces $\CH_N$ in the previous
section with the braided-symmetrized tensor products we have
obtained. Let the map $\pi: B_N \to S_N$ be the natural
homomorphism defined by identifying a crossing with its inverse
crossing. As in Section~2, we describe a topological configuration
$\tilde\gamma$ by the pair $(\gamma, b)$, where $p\,(\tilde\gamma)
= \gamma \in \Delta_N (\R^2)$, and where $b \in B_N$ (the
fundamental group of $\Delta_N$). Then with
$\gamma=\{x_1,\cdots,x_N\}$,
\[ \CH_N\,\,\isom\,\, \CH\tens_s\cdots\tens_s \CH,\quad
\tilde\Phi(\gamma,b)=f(\pi(b)(x_1,\dots,x_N)),\,\, \forall\,
x_1<x_2\cdots<x_N\,.\]
While $\gamma$ is unordered, we introduce the conventional
lexicographic ordering in indexing its elements.

For example if $x<y<z$ in $\mathbb{R}^2$, then
\[ \tilde\Phi(\{x,y,z\},e)=f(x,y,z),\quad
\tilde\Phi(\{x,y,z\},b_{23})=f(x,z,y)=\,q\,\tilde\Phi(\{x,y,z\},e)
=\,q\,f(x,y,z)\]
by equivariance, where $b_{23}$ is the braid group generator
braiding strand $2$ with strand $3$. In general, if
$x_1<\cdots<x_N$ then
\begin{equation}
 f(\sigma(x_1,\dots
,x_N))\,=\,\tilde\Phi(\{x_1,\dots,x_N\},i(\sigma))
\,=\,q^{\ell\,(\sigma)}\tilde\Phi(\{x_1,\dots,x_N\},e)
\,=\,q^{\ell\,(\sigma)}f(x_1,\dots,x_N)\,; \label{ingeneral}
\end{equation}
where $i(\sigma)$ is the braid defined as follows: Let $\sigma =
s_{j_1}\cdots s_{j_{\ell(\sigma)}}$ be a reduced expression for
$\sigma$ in terms of simple exchanges $s_j=(j,j+1)$. Let $b_j$ be
the braid group generator braiding strands$j$ and $j+1$. Then
$i(\sigma) = b_{i_1}\cdots b_{i_{\ell(\sigma)}}$. Note that
$i(\sigma)$ does not define a group homomorphism. Eq.
(\ref{ingeneral}) is just as required; our symmetry condition on
$f$ corresponds to the equivariance under $B_N$ of $\tilde\Phi$.

Likewise, using  the definitions in Section~2 and the diagrammatic
notation as in Ref. \cite{GS1996}, one obtains
\begin{eqnarray*}
(\psi^*(h)\tilde\Phi)(\{x,y,z\},e) &=&
q^{-\#(y,z<x)}h(x)\tilde\Phi(\{y,z\},e)\,+\,\,q^{-\#(x,z<y)}h(y)
\tilde\Phi(\{x,z\},e)\,+\\
& &\quad \quad \quad \quad \quad \quad \quad\,+\,\,
q^{-\#(x,y<z)}h(z)\tilde\Phi(\{x,y\},e)\,,
\end{eqnarray*}
as well as
\[(\psi(z)\tilde\Phi)(\{x,y\},e)=q^{\#(x,y<
z)}\tilde\Phi(\{x,y,z\},e)\,,\]
where $\#(x,y<z)$ denotes the number of points in the set
$\gamma=\{x,y\}$ to the left of $z$ in $\R^2$. These actions are
to be compared with (\ref{psih}) and (\ref{psistarh}). Similarly
one proves in general that the two constructions coincide; that
is, the representation of the anyon creation and annihilation
fields in the spaces $\CH_N$ of topological configurations is
equivalent to their representation in the braided Fock space by
$h\tens_s$ and the interior product.

\section{Roots of unity and the anyonic exclusion principle}

In this section we specialize further to the case where
$\,q=e^{2\pi\imath/r}\,$ is a primitive $r\,$th root of unity. Our
main observation is that in the continuum case, the explicit
representation (\ref{psi*h}) on $\Psi_0$-symmetric tensor products
implies that
\eqn{pauli}{ \psi(h)^r=0,\quad \psi^*(h)^r=0,\quad (\forall h)\,;}
that is, we have an {\em anyonic exclusion principle}. Here $h$ is
any test-function, and the smearing of $\psi$ by $h$ makes sense
of the products of distributions. By taking `bump functions' that
are increasingly localized at an arbitrary point $x$, we can also
write informally,
\eqn{paulix}{ \psi(x)^r=0,\quad \psi^*(x)^r=0,\quad (\forall x),}
a condition which we shall justify directly in the discrete case.
Before doing so let us note that conversely, the pointwise
condition (\ref{paulix}) in the discrete case implies
(\ref{pauli}), since
\begin{eqnarray}\label{linroot} \psi(\lambda\delta_x+\mu\delta_y)^r=
(\lambda\psi(x)+\mu\psi(y))^r &=&\sum_{s=0}^{s\,=\,r}
\lambda^s\mu^{r-s}\psi(x)^s\psi(y)^{r-s}\left[\,{r\atop
s}\,;q^{\eps_0(y,x)}\right]\nonumber\\ &=&
\lambda^r\psi(x)^r+\mu^r\psi(y)^r,\end{eqnarray}
for any constants $\lambda,\mu$ and $x\ne y$. In obtaining
(\ref{linroot}), we use the $q$-binomial theorem for $q$-commuting
quantities, with $q$-binomial coefficients defined in the usual
way but with $q$-integers in place of integers in the factorials.
Since $q^r=1$ we have $[r;q^{\pm1}]=0$ and hence only $s=0$ and
$s=r$ contribute. So the pointwise and smeared versions are
formally equivalent, with the smeared version being more suitable
in the continuum case.

To prove (\ref{pauli}) in the continuum, we use (\ref{psi*h})
acting on the vacuum (the identity function with no arguments) to
deduce that
\[
\psi^*(h)^m(x_1,\cdots,x_m)=[m;q^{-1}]!h(x_1)\cdots h(x_m)\] for
all non-negative integers $m$. This follows by induction on $m$.
It is true for $m = 1$; assuming it for $m$, we have
\begin{eqnarray*} \psi^*(h)^{m+1}(x_1,\cdots,x_{m+1})&=&
h(x_1)\psi^*(h)^m(x_2,\cdots,x_{m+1})
+q^{-1}h(x_2)\psi^*(h)^{m}(x_1,x_3,\cdots,x_{m+1})\\
&&+\cdots+q^{-m}h(x_{m+1})\psi^*(h)^m(x_1,\cdots,x_m)\\
&=&(1+q^{-1}+\cdots+q^{-m})[m;q^{-1}]h(x_1)\cdots
h(x_{m+1})\end{eqnarray*} as required. The exclusion principle
follows when $q^r=1$, since then $[r;q^{-1}]=0$ as already noted
above. Next, letting $\psi^*(h)^r$ act on any Fock state of the
form $\psi^*(h_1)\cdots\psi^*(h_N)|0\>$, we move $\psi^*(h)^r$ to
the right until it arrives to act on the vacuum. In doing so, $h$
will be replaced by a convolution with all the $h_1,\cdots h_N$
[see Eqs. (\ref{q-weylh})], giving us a new test function $h'$.
But we have already verified that $\psi^*(h')^r|0\>$ for all test
functions; hence (\ref{pauli}) holds in the representation. The
result for $\psi(h)^r$ follows, as in the continuum case it is the
adjoint.

In fact, in the continuum theory one can formally conclude
(\ref{pauli}) as a statement about operator-valued distributions.
We exhibit the reasoning for $r=3$ (the general case is similar).
We have
\begin{eqnarray*} \psi^*(h)^3&=&\int dx dy dz\,
h(x)h(y)h(z)\psi^*(x)\psi^*(y)\psi^*(z)\\
&=&\sum_{\sigma\in S_3}\int_{\sigma(x)<\sigma(y)<\sigma(z)}dx dy
dz\, h(x)h(y)h(z)
q^{-\ell(\sigma)}\psi^*(\sigma(x))\psi^*(\sigma(y))\psi^*(\sigma(z))\\
&=&\sum_{\sigma\in S_3}q^{-\ell(\sigma)}\int_{x<y<z}dx dy dz\,
\psi^*(x)\psi^*(y)\psi^*(z)=0\,,\end{eqnarray*}
where we have written
$\sigma(x,y,z)=(\sigma(x),\sigma(y),\sigma(z))$, and where
$\ell(\sigma)$ is the length of the permutation. In effect we have
broken up the $3$-fold integral into $3!$ permutations of the
fundamental domain where $x<y<z$ (not being concerned about
coincident points since these form a subset of measure zero); then
we have used the $q$-commutation relations (\ref{q-weyl}), and
finally we have changed variables to give many copies of the
integral over the fundamental domain. The result is zero since
$\sum q^{-\ell(\sigma)}=[3;q^{-1}]!=0$ when $q^3=1$. The same
argument holds for general $r$ and for $\psi(h)^r=0$.

The arguments for the discrete case are more subtle, and indeed
the formulas $\psi(x)^r=0$ and $\psi^*(x)^r=0$ do not hold
automatically in our representation. Rather, we argue that in this
case it is natural to {\it impose\/} these according to the
algebraic structure. This is similar to `truncated versions' of
quantum groups and other $q$-algebras in conformal field theory
and other settings at roots of unity. Indeed we then have
$\Psi^r=\id$ for the braiding, and we are essentially in the
setting that has been called `anyonic vector spaces' in
\cite{Ma:any} and \cite{Ma:book}.

{}From a physical point of view, the key observation is that for
$q$ a primitive $r\,$th root of unity, the operators
\[ \psi(x)^r,\quad \psi^*(x)^r\]
are in any case central. In fact, from (\ref{q-weyl}) we have
$\psi(x)^m\psi(y)=q^{m\eps_0(x,y)}\psi(y)\psi(x)^m$, and
\begin{eqnarray*} \psi(x)^m\psi^*(y)&=&\psi(x)^{m-1}q^{\eps(y,x)}
\psi^*(y)\psi(x)+\psi(x)^{m-1}\delta(x-y)\\
&=&
q^{m\eps(y,x)}\psi^*(y)\psi(x)^m+\delta(x-y)\psi(x)^{m-1}[m;q^{\eps(y,x)}].\end{eqnarray*}
Hence, when $q^r=1$, we have $[r\,;q^{\pm1}]=0$ and $\psi(x)^r$ is
central. Note that here it is critical that we used $\eps$ and not
$\eps_0$ in the calculation involving $\psi^*(y)$. Similarly, we
have that $\psi^*(x)^r$ central. Therefore, in an irreducible
sector of the theory, these operators should be set to multiples
of the identity.

This argument is needed only in the discrete case, but the
algebraic structures apply (suitably understood) in both cases;
thus we have used a notation applicable to both.

As for which value these operators should be assigned, we have
already seen that (\ref{paulix}) is suitable for the constraint to
be linear (basis independent), in the sense of applying to all
$h$. Clearly zero is the only value with this property. Another
way to reach the same conclusion is in terms of the braided
coproduct on $S_{\Psi_0}(\CH)$. On products it extends by $\Psi$,
and one has
\[ \und\Delta \psi^*(x)^r= \sum_{s=0}^{s=r}\psi^*(x)^s\tens
\psi^*(x)^{r-s} \left[{r\atop s};q\right]= \psi^*(x)^r\tens 1
+1\tens \psi^*(x)^r.\]
A similar result holds for $\psi(x)^r$. Since $\und\Delta 1=1\tens
1$, only (\ref{paulix}) allows the braided coproduct to descend to
the reduced algebra; no other constant will do. Moreover, since
this $\und\Delta$ underlies the braided-differentiation operation
$\del^x$ in Section~3.1, our representation of $\psi^*$ (and more
obviously $\psi$) then descends to an action of the truncated
algebra $S_{\Psi_0}(\CH)$ (in which (\ref{paulix}) is imposed) on
itself, by the same formulas as before. This is in subtle contrast
to the continuum case, where (\ref{pauli}) already holds.

Next we note that because  $\psi^*(x)$ and $\psi^*(y)$ commute up
to a factor $q^{\eps_0(x,y)}$, any state formed by a sequence of
creation field operators applied to the vacuum, where $r$ or more
occurrences of $\psi^*(x)$ are included, must vanish. That is,
\[ \psi^*(x)\psi^*(y)\psi^*(z)\cdots \psi^*(x)\psi^*(w)
\cdots\psi^*(x)\cdots |\,0\,\>=0\]
if there are $r$ or more instances of $\psi^*(x)$. The distinct
$\psi^*(y)$, $\psi^*(z)$ and so forth can equally be annihilation
fields $\psi(y),\psi(z)$ and so forth, since these too $q$-commute
with $\psi^*(x)$.

More generally, a state formed from the vacuum will be zero if
there are $m$ annihilation $\psi(x)$ operators, and $r+m$ or more
$\psi^*(x)$. This can be proved by induction on $m$, as follows.
The case $m=0$ is covered above. Suppose it is true for $m-1$.
Given an expression with $m$ annihilation operators, look at the
rightmost $\psi(x)$. Applying the $q$-commutation relation with
the $\psi^*$ to its right, we move $\psi(x)$ to the right and pick
up a second term with a $\delta$-function and with one fewer
$\psi(x)$, and at most one fewer $\psi^*(x)$; by our induction
hypothesis, this second term vanishes. Meanwhile, the first term
has $\psi(x)$ one step to the right; repeating this eventually
brings it to act directly on $|\,0\,\>$, giving zero. This proof
makes sense in the discrete case, or with the assumption that the
$\delta$-functions are approximated by bounded functions with the
limit taken only at the end (in order to treat $\delta(0)$ as a
number). More precisely, in terms of annihilation operators
smeared with test functions,
\[ \psi^*(x)\cdots \psi(h_1)\cdots \psi^*(x) \cdots
\psi(h_m)\cdots \psi^*(x)\cdots |\,0\,\>=0\]
if there are $m$ annihilation operators and at least $r+m$
creations $\psi^*(x)$ anywhere in the string. This follows from
(\ref{q-weylh}) and a similar proof by induction.

For general $\psi^*(h)$ we always have the exclusion principle
(\ref{pauli}). But the stronger version, in which some of the
$\psi^*(h)$ are not adjacent, is more complicated. The various
$\psi^*(h)$ needed are modified by the intervening creation field
operators, according to (\ref{q-weylh}). Thus, one has instead
exclusion conditions that take the form
\[
\psi^*(\theta_0{}_{x_1}\cdots\theta_0{}_{x_m} h)^{n_0}
\psi^*(x_1)\psi^*(\theta_0{}_{x_2}\cdots\theta_0{}_{x_m}
h)^{n_1}\cdots\quad\quad
\]
\[
\quad\quad\cdots\psi^*(x_{m-1})\psi^*(\theta_0{}_{x_m}h)^{n_{m-1}}
\psi^*(x_m)\psi^*(h)^{n_m}\,|\,0\,\>\,=\,0,
\]
when $n_0+\cdots+n_m\,\ge\, r$. In the Bose or Fermi cases, the
functions $\theta_0{}_x$ are constant ($\pm 1$), up to a set of
measure zero; but otherwise they must be taken into account. The
same complication applies when there are annihilation operators
present. This would appear to be a feature of the anyonic theory
($r>2$), that is not present for fermions.

On the other hand, we do not see this complication if all our
smeared fields have disjoint support. Thus, from (\ref{q-weylh})
we find that \[
\psi^*(h)\psi^*(g)=\psi^*(g)\psi^*(h)\begin{cases}q & {\rm if} \
{\rm supp}(h)<{\rm supp}(g)\\ q^{-1}& {\rm if}\ {\rm supp}(h)>{\rm
supp}(g)\end{cases}\] \[ \psi(h)\psi^*(g)=(\bar
g,h)+\psi^*(g)\psi^*(h)\begin{cases}q^{-1} & {\rm if} \ {\rm
supp}(h)<{\rm supp}(g)\\ q& {\rm if}\ {\rm supp}(h)>{\rm
supp}(g)\end{cases}\] with a similar equation for
$\psi(h)\psi(g)$. When they occur, these have a similar form to
(\ref{q-weyl}). Hence, as an application, one may take the
$\psi^*(x)$ as given more precisely by smearing with `bump
functions' of small support around the relevant point. As long as
these bumps to not touch, the various smeared $\psi^*(x)$ fields
behave as in the discrete case, and may be collected together by
similar relations. For such states we have the full exclusion
principle again, without any complications when the instances of
the smeared field $\psi^*(x)$ are separated from each other in the
product of field operators.

With this last observation in mind, let us give a straightforward
application of the exclusion principle to a gas of non-interacting
anyonic particles (e.g., vortices) localized in disjoint sets
around points $x_1,\cdots, x_n$ in $\R^2$. Let us suppose that
each particle carries a fixed unit $E$ of energy, which does not
depend on the positions; the latter will therefore be considered
as fixed (or else, we must factor out the resulting degeneracy).
As per the discrete version of the theory, the reduced range of
states is then:
\[ |m_1,\cdots,m_n\>=\psi^*(x_1)^{m_1}\cdots
\psi^*(x_n)^{m_n}|\,0\>,\quad m_i=0,\cdots,r-1.\]
Let us decompose this reduced Hilbert space as $\,\oplus_N
\CH_N\,$ according to $N \,=\,\sum_{i=1}^n m_i$, which is the
value of the occupation number operator
\[ {\mathcal{N}}=\int dx\, \rho(x)=\int dx\, \psi^*(x)\psi(x).\]
The statistical partition function is then
\eqn{Zbeta}{ Z_\beta={\rm Trace}(e^{-\beta E
{\mathcal{N}}})=\sum_{N}e^{-\beta E N}\dim(\CH_N)=\sum_{\{m_i\}}
e^{-\beta E\sum_i m_i}=[r;e^{-\beta E}]^n\,.}
The $q$-integer again appears, but now at the real value
$e^{-\beta E}$. Each mode $\psi^*_i$ may be counted separately, so
that the computation here is the same as that for $1$ particle,
but raised to the $n$th power. When $r=2$ we recover the usual
partition function for fermions, while for $r=\infty$ we recover
the usual result for bosons:
\[ [2;e^{-\beta E}]=e^{-\beta E}+1;\quad [\infty;e^{-\beta E}]=
{1\over 1-e^{-\beta E}}\,.\]
The general formula $[r;e^{-\beta E}]$ interpolates the two. From
the partition function one may then proceed as usual to
thermodynamic properties of such a gas \cite{AcharyaSwami1994}.

More generally, we may take Hamiltonians with interaction terms,
including those that arise from the field theory rather than the
harmonic oscillator point of view (i.e., with kinetic and
current-current interaction terms). Such applications will be
developed elsewhere.

\section{Concluding remarks}

We have obtained a generalized exclusion principle for anyons---an
important physical result---not from analysis of the statistical
mechanics of anyons, but from the explicit representation of
nonrelativistic creation and annihilation fields in an appropriate
braided tensor product space. As one would expect in this context,
the principle holds when the anyonic phase shift is a root of
unity. It applies both to smeared fields $\psi(h),\psi^*(h)$ and
to (unsmeared) operator-valued distributions $\psi(x),\psi^*(x)$
satisfying $q$-commutation relations; but it takes a rather
cleaner form in the latter case (a subtle distinction that
disappears for bosons and fermions).

On the other hand, the discrete version in which we work directly
with points (rather then with increasingly peaked bump functions)
turns out to be different and algebraically more complicated; with
$\psi^i$ and $\psi^*_i$ no longer adjoint to each other. The fact
that one has a different theory from the continuum limit is an
interesting feature of our analysis. We believe this subject, and
its relation to Gentile statistics and to generalized harmonic
oscillators, deserves some renewed attention.

Let us also mention some related conceptual aspects of interest.
Our construction of $S_{\Psi_0}(\CH)$ in Section~2 is manifestly
dependent on an ordering, since this enters in the braiding. This
is true for quantum planes (see the remarks at the start of
Section~5), where $X=\{1,2,\cdots,n\}$ is the indexing set; and it
remains true when we apply our formalism to the second
quantization of nonrelativistic fields (so that $X$ denotes
physical space). The Fock space construction might then seem to be
strongly dependent on the somewhat artificial lexicographical
total ordering on $\R^2$ used in Section~3. However the
isomorphism in Section~4, with the spaces $\CH_N$ described by
means of topological configurations, tells us that in fact the
underlying diffeomorphism invariance remains as far as the physics
is concerned. The lexicographical ordering places the physical
system into an algebraic form described by the symmetric tensor
products $\tens_s$, but this is for mathematical convenience only.

This suggests an answer to a certain puzzle in $q$-deformed
physics---how to physically interpret noncommutative tensor
products (as generated by non-cocommutative quantum groups). That
is, if $A$ and $B$ are two physically equivalent systems, what is
the difference between $A\tens_s B$ and $B\tens_s A$, and which is
the correct description of the joint system? In our anyonic model
this is an unphysical distinction that is needed to work
algebraically; just as (physically) the points
$\gamma=\{x_1,\cdots,x_N\}$ in $\Delta_N$ are intrinsically
unordered, but it can nevertheless be helpful (mathematically) to
order them. This is the difference between diffeomorphisms of the
manifold $\R^2$ acting on subsets of $\R^2$, and the coordinate
description of their lifting to the universal covering space of
$N$-identical-particle configuration space.

Another interesting feature is the way that the strictness of the
braiding $\Psi$ comes about from the structure of the
singularities at coincident points (expressed here as Kronecker or
Dirac $\delta$-functions). In general one has diagonal
singularities when multiplying operator fields, but in some cases
(such as in conformal field theory) these can be controlled (e.g.,
by the operator product expansion). It would be interesting to see
how braidings that arise in conformal field theory relate to
diffeomorphism group ideas and to $q$-Fock space ideas along the
lines of the present paper. In some situations, such as the
Wess-Zumino-Witten model, there is also a topological path picture
leading to the quantum group $U_q(su_2)$.

Finally, we note that the basic ideas described here apply also to
plektons---particles associated with higher-dimensional
(non-Abelian) unitary representations of the braid group. Here
$T(b)$ in Section~2 is not simply a phase, but a
finite-dimensional unitary operator acting on a multicomponent
wave function; and we work with a multiplet of operator fields.
Essentially, we then have an $R$-matrix for the linear braid group
representation, in place of $q$ in the formulas above; and we must
also be more careful about ordering. Thus the creation and
annihilation fields act no longer by multiplication by powers of
$q$ representing the number of crossings in the resulting braid,
but matrix operation on the multiplet. In place of $q^{\eps(x,y)}$
and $q^{\eps_0(x,y)}$, we define  $\Psi$ and $\Psi_0$, using $R$
or $R^{-1}$ according to the ordering---but with the same
conceptual picture that we have used in the anyonic case.

\bigskip

\baselineskip 14pt
\end{document}